\documentclass[twocolumn,preprintnumbers,amsmath,amssymb]{revtex4}
\usepackage{color}
\usepackage{graphicx}
\usepackage{dcolumn}
\usepackage{bm}

\begin{document}


\title{Magnetic and structural transitions in layered FeAs systems:\\ AFe$_2$As$_2$ versus RFeAsO compounds}

\author{C. Krellner}
\author{N. Caroca-Canales}
\author{A. Jesche}
\author{H. Rosner}
\author{A. Ormeci }
\author{C. Geibel}
\email{geibel@cpfs.mpg.de}
\affiliation{Max Planck Institute for Chemical Physics of Solids, N\"othnitzer Str. 40, D-01187 Dresden, Germany }

\date{\today}

\begin{abstract}
Resistivity, specific heat and magnetic susceptibility measurements performed on SrFe$_2$As$_2$ samples evidence a behavior very similar to that observed in LaFeAsO and BaFe$_2$As$_2$ with the difference that the formation of the SDW and the lattice deformation occur in a pronounced first order transition at $T_0=205$\,K. Comparing further data evidences that the Fe-magnetism is stronger in SrFe$_2$As$_2$ and in EuFe$_2$As$_2$ than in the other layered FeAs systems investigated up to now. Full potential LDA band structure calculations confirm the large similarity between the compounds, especially for the relevant low energy Fe $3d$ states. The relation between structural details and magnetic order is analyzed.
\end{abstract}

                             

\maketitle

\begin{figure}[t]
\includegraphics[width=8.5cm]{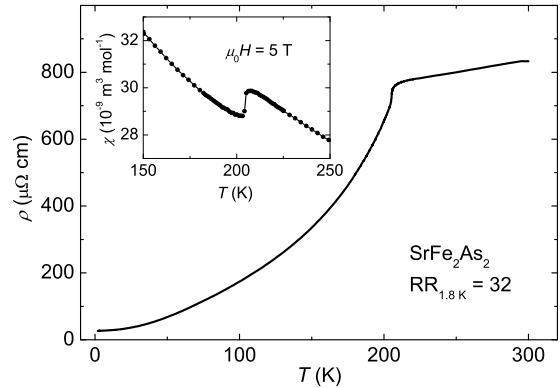}
 \caption{\label{FigRvT} Temperature dependence of the electrical resistivity $\rho(T)$ of SrFe$_2$As$_2$. A step-like decrease marks the first order phase transition at $T_0=205$\,K. The magnetic susceptibility $\chi(T)$ measured at $\mu_0H=5$\,T is shown in the inset.}
\end{figure}

\begin{figure}[t]
\includegraphics[width=8.5cm]{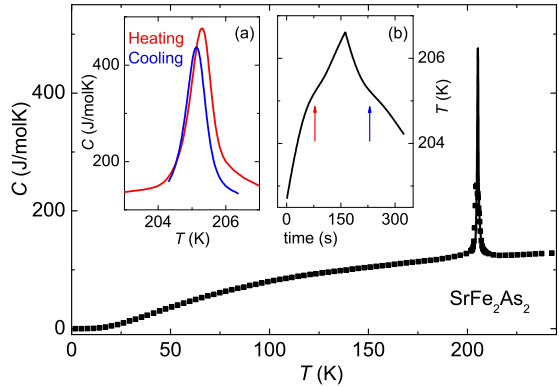}
  \caption{\label{FigCvT} (Color online) Specific heat of SrFe$_2$As$_2$ plotted as $C$ vs $T$. A sharp first order type anomaly is visible at $T_0=205$\,K. This anomaly is enlarged in inset (a), determined from a heating and a cooling curve in a relaxation-time measurement, shown in inset (b). The arrests in the temperature-time relaxation curve due to latent heat are depicted by arrows.}
\end{figure}

The discovery of superconductivity in doped LaFeAsO \cite{Kamihara:2008} and the subsequent raising of the superconducting (SC) transition temperature $T_c$ to 56\,K \cite{Takahashi:2008, Ren:2008, Chen:2008} initiated a surge of interest in layered FeAs systems. Undoped RFeAsO compounds (R = La - Gd) present a structural transition at $T_0 \sim 150$\,K followed by the formation of a spin density wave (SDW) at a slightly lower temperature $T_N \sim 140$\,K \cite{Cruz:2008, Klauss:2008}. Electron or hole doping by substituting O by F \cite{ChenNAture:2008}, or R by a tetravalent or divalent cation \cite{Wen:2008, Wang:2008}, or by reducing the O-content \cite{Ren:2008a} leads to the suppression of the SDW and to the onset of superconductivity. This connection between a vanishing magnetic transition and the simultaneous formation of a SC state is reminiscent of the behavior in the cuprates and in the heavy fermion systems, and therefore suggests the SC state in these doped RFeAsO systems to be of unconventional nature, too. While this has to be confirmed by further studies, there seems to be a general belief that the intriguing properties of these compounds are connected with very peculiar properties of the FeAs layers.
\\
The tetragonal ZrCuSiAs structure type in which these compounds crystallize \cite{Quebe:2000} results in a square Fe lattice with As in the center of the square but being alternately shifted above and below the Fe-plane. It is well known that the ThCr$_2$Si$_2$ structure type presents a very similar arrangement of the transition metal and $p$-element. Thus it was natural to look for appropriate candidates within the huge amount of compounds crystallizing in this structure type. In a very primitive approach, the RFeAsO compounds can be rationalized as a stack of alternating (Fe$_2$As$_2$)$^{2-}$ and (R$_2$O$_2$)$^{2+}$ layers. Exchanging the (R$_2$O$_2$)$^{2+}$ layer by a layer with a single large atom A leads to the ThCr$_2$Si$_2$ structure type. In order to keep the same electron counts as in the RFeAsO materials, A has to be a divalent atom. Therefore, appropriate candidates are A$^{2+}$Fe$_2$As$_2$ compounds. To the best of our knowledge, three of the possible candidates have already been investigated. M. Pfisterer and G. Nagorsen have reported the synthesis, the structure as well as preliminary susceptibility data of SrFe$_2$As$_2$ and BaFe$_2$As$_2$ \cite{Pfisterer:1980, Pfisterer:1983}.  From their susceptibility data they concluded the occurrence of a magnetic phase transition around 200\,K and 130\,K, respectively and suggested it to be an antiferromagnetic (AFM) one. Later on, Raffius \textit{et al.} presented a more detailed investigation of EuFe$_2$As$_2$, which unambiguously demonstrated AFM ordering of Fe at 200\,K (as well as AFM ordering of Eu$^{2+}$ at 19\,K) \cite{Raffius:1993}. From the analysis of the M\"ossbauer data, which revealed e.g. a very small hyperfine field, they concluded the Fe magnetism to be itinerant. Interestingly, they reported in the same paper temperature dependent M\"ossbauer spectra of LaFeAs, which was later corrected to be LaFeAsO and infered that in this compound Fe also presents itinerant magnetism and orders magnetically at 139\,K \cite{LaFeAsO}. Thus, the presence of itinerant magnetism in LaFeAsO as well as the magnetic ordering at around 134\,K is not an original discovery in the very recent publications but was already established by a microscopic method 15 years ago.
\\
In the context of the present paper, it is more important to notice that the Fe-M\"ossbauer data in EuFe$_2$As$_2$ (where Eu was proven to be divalent) are quite similar to those in LaFeAsO, with only a slightly larger (but still very small) hyperfine field. This similarity, as well as some resemblance between the susceptibility data of SrFe$_2$As$_2$ with those of LaFeAsO gave a first evidence for a strong analogy between the magnetism in A$^{2+}$Fe$_2$As$_2$ and that in LaFeAsO. For revealing and understanding the origin of anomalous behavior, comparing and analyzing the properties of closely related compounds showing this behavior in slightly different strength is one of the most successful approaches in condensed matter physics. In order to start this process, we present here an experimental study of the basic magnetic properties of SrFe$_2$As$_2$ combined with LDA based band structure calculations. While some of the results indeed confirm the closeness of both types of compounds, a more detailed analysis revealed significant differences, which may provide a way to discriminate between different proposed theoretical models. While finalizing our investigation Rotter \textit{et al.} submitted a paper on BaFe$_2$As$_2$ \cite{Rotter:2008}. They confirm the presence of a second order phase transition at 140\,K where both SDW and lattice distortion set in. The properties they reported for BaFe$_2$As$_2$ are very similar to those reported for the RFeAsO compounds.\\
The polycrystalline samples of SrFe$_2$As$_2$ were synthesized by heating a stoichiometric mixture of Sr (3.5N), Fe (4N), and As (7N) in an Al$_2$O$_3$ crucible, sealed under an inert atmosphere (Ar 6.0) inside an evacuated quartz tube. In a first step, the elements were heated very slowly to 950$^{\circ}$C and kept at this temperature for 10 hours; subsequently, the mixture was cooled down to room temperature. The reaction product was ground and pelletized and for a second heating step sealed inside a Ta-crucible and sintered at 1100$^{\circ}$C, again for 10 hours. The resulting pellet with a metallic luster was found to be unstable in air and therefore, was kept under an inert atmosphere. Powder of this pellet was characterized by X-ray diffraction (XRD) recorded on a \textsc{Stoe} diffractometer in transmission mode using monochromatic Cu-K$_{\alpha}$ radiation ($\lambda = 1.5406$\,\AA). The resulting XRD pattern can be well indexed on the basis of a tetragonal ThCr$_2$Si$_2$-structure type  and did not reveal any foreign phase peak. The lattice parameters $a = 3.924(3)$\,\AA\, and $c = 12.38(1)$\,\AA\, refined by simple least square fitting were found to be in good agreement with the reported structure data \cite{Pfisterer:1980}. The DC-susceptibility $\chi(T)$ was measured in a commercial Quantum Design (QD) magnetic property measurement system (MPMS). AC-resistivity  $\rho(T)$ measurements down to 1.8\,K were carried out in a standard four-probe geometry using a QD physical property measurement system (PPMS). The PPMS was also used to determine the specific heat $C(T)$ with a standard heat-pulse relaxation technique. To gain deeper insight into the electronic structures of the investigated compound on a microscopic level, we performed density functional band structure calculations within the local (spin) density approximation (L(S)DA). Using the experimental structural parameters \cite{Cruz:2008, Pfisterer:1980, Pfisterer:1983, Rotter:2008}, we applied the full-potential local-orbital code (FPLO) \cite{Koepernik:1999} (version 7.00-28) with the Perdew-Wang exchange correlation potential \cite{Perdew:1992} and a well converged $k$-mesh of 24$^3$ points for the Brillouin zone. 
\\
The most prominent features are observed in the resistivity and in the specific heat, while the temperature dependence of the susceptibility is rather weak. Below 300\,K, $\rho(T)$ decreases slightly with temperature $T$ (Fig.~\ref{FigRvT}) and thus evidences metallic behavior, although the absolute value at 300\,K is somewhat larger than in a classical metal (but well below the values observed in typical LaFeAsO samples) and the slope $\partial \rho(T)/\partial T$ is comparatively small. At $T_0=205$\,K, we observe a sharp step-like drop in $\rho(T)$ with a step size $\Delta \rho/\rho_{210\,\rm K}\sim 6$\%, followed by a continuous decrease by more than one order of magnitude down to 1.8\,K. This results in a resistivity ratio (RR) of about $\rho_{300\,\rm K}/\rho_{1.8\,\rm K}=32$, which is much larger than that observed in LaFeAsO \cite{Klauss:2008} or in BaFe$_2$As$_2$ \cite{Rotter:2008}. Since such a strong decrease can hardly be attributed to the vanishing of spin disorder scattering, it indicates pronounced changes of the electronic states at the Fermi level.
\\
In contrast, $\chi(T)$ increases only slightly with decreasing temperature below 300\,K (inset of Fig.~\ref{FigRvT}), the increase being more pronounced at small applied fields which indicates this increase to be due to paramagnetic impurities or magnetic foreign phases. The susceptibility above $T_0$ is compatible with an enhanced Pauli susceptibility. At $T_0$, $\chi(T)$ also presents a step-like decrease, but there is no further decrease at lower temperatures. The size of the step at $T_0$, $\Delta \chi = 1.1\cdot 10^{-9}$ m$^3$/mol corresponds to a relative change $\Delta \chi/\chi_{210\,\rm K}\sim 5$\%, quite similar to the relative size of the step in $\rho(T)$ of 6\%. In an itinerant scenario, that would suggest a decrease of the density of states at the Fermi level by just 5\%. The specific heat $C(T)$ (Fig.~\ref{FigCvT}) provides a very clear evidence for a pronounced first order transition at $T_0$. On the top of a monotonously with $T$ increasing $C(T)$, as expected from the phonon contribution, we found a very sharp and huge peak centered at $T_0=205$\,K with a half width at half maximum of less than 0.4\,K. For tracing such a sharp peak, instead of determining a single $C(T)$ value by fitting the whole relaxation curve with an exponential function, it is more appropriate to calculate a continuous $C(T)$ curve from the time derivative of the relaxation curve directly \cite{Lashley:2003}. The relaxation curve across the transition is shown in the right inset of Fig.~\ref{FigCvT}, while $C(T)$ determined directly from the slope is shown in the left inset. One immediately notices the shoulders (arrows) in the relaxation curve which are a direct evidence for a thermal arrest and thus for a first order transition (see e.g. \cite{Lashley:2003}). Also the hysteresis between $C(T)$ determined from the cooling and the heating part of the relaxation curve indicate a first order transition. From the area under the peak in $C(T)$, one can estimate the latent heat $\Delta H$ as well as the entropy $\Delta S$ connected with the transition. They amount to $\Delta H \sim  200$\,J/mol and $\Delta S\sim 1$\,J/molK. This small value of $\Delta S$ can be taken as evidence for an itinerant Fe moment. Muon spin relaxation and XRD measurements confirm that the formation of the SDW and the lattice deformation occurs simultaneously at $T_0$, which will be presented in a forthcoming paper \cite{Jesche:2008}.
\\
Our experimental results confirm the presence of a single transition at $T_0=205$\,K in SrFe$_2$As$_2$ and show that the overall behavior of the susceptibility and of the resistivity of SrFe$_2$As$_2$ is quite similar to that of the undoped RFeAsO compounds and BaFe$_2$As$_2$. This suggests that the underlying physics is the same. However, a more detailed comparison indicates that BaFe$_2$As$_2$ is almost identical to the RFeAsO series, while the replacement of Ba by the smaller Sr or Eu$^{2+}$ enhances the magnetic character of the FeAs layers. This is obvious from the ordering temperature which increases from $T_0\sim150$\,K in the former ones to $T_0\sim200$\,K in the latter ones. EuFe$_2$As$_2$ and SrFe$_2$As$_2$ being almost identical is not surprising, since Sr has almost the same ionic radius as Eu$^{2+}$ and is often used as a non magnetic reference for Eu$^{2+}$ compounds. The increase of the ordering temperatures is accompanied by an increase of the ordered Fe moment as evidenced by the increase of the hyperfine field $B_{\rm hf}$ determined in Fe-M\"ossbauer experiments from $B_{\rm hf}\sim5.1\pm0.2$\,T in LaFeAsO \cite{Raffius:1993, Kitao:2008} and BaFe$_2$As$_2$ \cite{Rotter:2008} to $B_{\rm hf} = 8.5$\,T in EuFe$_2$As$_2$ \cite{Raffius:1993}. Furthermore, the value of the Pauli-like susceptibility above the transition also increases from $\sim 5\cdot 10^{-9}$\,m$^3$/mol-Fe in LaFeAsO \cite{Klauss:2008} and $\sim 6\cdot 10^{-9}$\,m$^3$/mol-Fe in BaFe$_2$As$_2$ \cite{Rotter:2008} to $\sim 13\cdot 10^{-9}$\,m$^3$/mol-Fe in SrFe$_2$As$_2$. This increase in the magnetic character of EuFe$_2$As$_2$ and SrFe$_2$As$_2$ cannot be easily related to a change in the structural parameters. The lattice parameter $a$ decreases from BaFe$_2$As$_2$ ($a=3.9625$\,\AA, \cite{Rotter:2008})  to SrFe$_2$As$_2$ ($a=3.924$\,\AA) and EuFe$_2$As$_2$ ($a=3.911$\,\AA \cite{Marchand:1978}), but this reduction is smaller than that within the RFeAsO series. There $a$ decreases from 4.0355\,\AA\, in LaFeAsO \cite{Kamihara:2008} to 3.9151\,\AA\, in GdFeAsO \cite{Wang:2008}, without increase in $T_0$. Also the buckling of the FeAs layers does not seem to change strongly as deduced from the distance of the As atoms to the Fe plane, which within the present accuracy of the reported data stays around $1.40\pm0.05$\,\AA. Thus the increase of $T_0$ by 35\% in SrFe$_2$As$_2$ and EuFe$_2$As$_2$ compared to the other compounds has to be related to more subtle changes in the electronic properties.
\begin{figure}[t]
\includegraphics[height=10cm, angle=270]{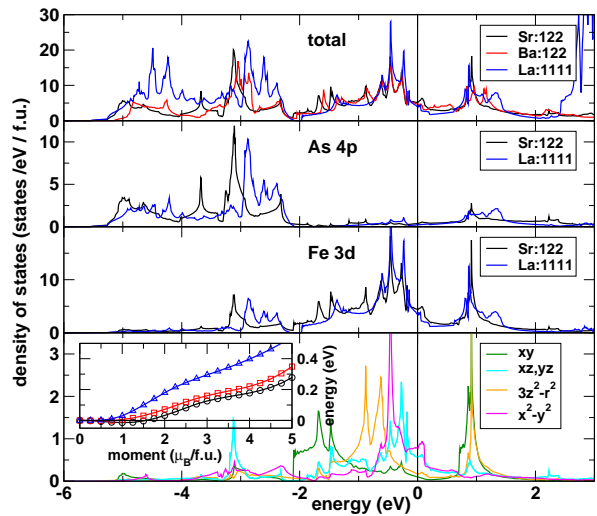}
\caption{\label{dos} (Color online) Total and partial DOS for  SrFe$_2$As$_2$ (Sr:122), BaFe$_2$As$_2$ (Ba:122) and LaFeAsO (La:1111). The bottom panel shows the orbital resolved Fe 3$d$ DOS for SrFe$_2$As$_2$. The inset provides the FSM results for the three compounds, the symbols present the calculated points.}
\end{figure}
\\
The calculated electronic density of states (DOS) for LaFeAsO and AFe$_2$As$_2$ (A=Sr, Ba) is shown in Fig.~\ref{dos}. On a first glance, the total DOS of the AFe$_2$As$_2$ systems is basically identical, evidencing the strong analogy between the two compounds even on small energy scales. However, also the LaFeAsO system exhibits a very similar electronic structure, especially in the relevant low energy region that is dominated by Fe 3$d$ states for all three compounds. The main differences appear in the region below 4\,eV due to O 2$p$ states and above 2.5\,eV due to the La 4$f$ states that are naturally absent in the other two systems. Smaller differences between the AFe$_2$As$_2$ compounds and the LaFeAsO system are observed in the region between -4\,eV and -2\,eV, where the main contribution is due to the As 4$p$ -- Fe 3$d$ hybrid states. For LaFeAsO, these states are shifted to higher energies by about 300\,meV (see Fig.~\ref{dos}). A surprisingly close similarity of all three systems is observed for the Fe 3$d$ states close to the Fermi level, where these states provide the predominant contribution to the total DOS. This close similarity even holds for the orbital resolved Fe 3$d$ DOS (shown for SrFe$_2$As$_2$ in the bottom panel of Fig.~\ref{dos}). These results suggest that all three systems should behave very alike with respect to their physical properties, including the response upon doping, pressure and other external parameters.
\\
However, a closer analysis of the orbital resolved Fe\,3$d$ DOS yields a strong dependency of the occupation of the Fe 3$d_{x^2-y^2}$ orbital on structural details like the exact As $z$ position. These small differences influence crucially the magnetic behavior: Whereas we find a magnetic ground state (GS) for all three compounds using the experimental As position, we find a nonmagnetic state with optimized As $z$ parameters \footnote{using the experimental lattice parameters} which results in slightly smaller ($\sim 0.07$\,\AA ) Fe-As distances.  The occupation of the Fe\,3$d_{x^2-y^2}$ orbital increases with decreasing Fe-As distances because of a downshift of the related band near $\Gamma$. This shift is connected to a considerable number of states because this band is almost dispersionless along $[001]$. If this band falls below the Fermi level along $[001]$, the magnetic state breaks down. 
\\
For the experimental structure, we find an antiferromagnetic GS for all three systems with a Fe moment of about 1.7\,$\mu_B$. Analyzing fixed spin moment (FSM) calculations (inset Fig.~\ref{dos}), we also find evidence for a ferromagnetic (FM) instability from minima in the FSM curves. These minima (m$_{\rm Sr}$=1.02\,$\mu_B$, m$_{\rm Ba}$=0.50\,$\mu_B$, m$_{\rm La}$=0.25\,$\mu_B$) coincide with the FM moments obtained in unconstrained calculations. The larger size of the magnetic moment and the energy dependence of the FSM curve support the experimental observation of stronger magnetic character of SrFe$_2$As$_2$. Interestingly, for all compounds the FSM curves show a second anomaly as a reminiscence of the AFM state. In addition, we investigated the structural stability with respect to the orthorhombic splitting for BaFe$_2$As$_2$ and SrFe$_2$As$_2$ (using the splitting ratio reported for BaFe$_2$As$_2$ \cite{Rotter:2008}) calculating the respective total energies. For both compounds, the orthorhombic solution is more stable. In the case of SrFe$_2$As$_2$, the calculated energy difference of about 300\,J/mol (370\,J/mol for BaFe$_2$As$_2$) compares quite well with the measured latent heat $\Delta H$ of 200\,J/mol.
\\
In summary, we prepared polycrystalline SrFe$_2$As$_2$ samples and measured the resistivity, magnetic susceptibility and specific heat. All the results evidence a very pronounced first order transition at $T_0=205$\,K with a width of less than 1\,K, in contrast to the broader, second order like transitions observed in the RFeAsO series and in BaFe$_2$As$_2$. The resistivity decreases strongly below $T_0$ leading to a resistivity ratio of 32 at 1.8\,K, which is much larger than that found in the other compounds. In contrast, the susceptibility is almost $T$ independent and presents only a small step at $T_0$. This behavior is very similar to that reported for the RFeAsO compounds and for BaFe$_2$As$_2$, which indicates that the underlying physics is the same. However, the larger ordering temperature in SrFe$_2$As$_2$ and EuFe$_2$As$_2$, the larger value of the Pauli like susceptibility above $T_0$ in SrFe$_2$As$_2$ as well as the larger Fe hyperfine field observed in M\"ossbauer experiments indicate a stronger magnetic character in these two compounds. The LDA band structure calculations evidence a pronounced similarity between the RFeAsO and the A$^{2+}$Fe$_2$As$_2$ compounds, especially for the relevant low energy Fe $3d$ states. They confirm the magnetic order to be more stable in SrFe$_2$As$_2$ than in BaFe$_2$As$_2$ or in LaFeAsO. A closer analysis reveals a strong dependence of the occupation of the Fe $3d_{x^2-y^2}$ orbitals on structural details like the exact As $z$ position, with a large impact on the Fe magnetism. Because of its stronger magnetic character including a larger $T_0$, SrFe$_2$As$_2$ likely offers a unique possibility for assessing the relevance of magnetic interactions for the superconductivity in layered FeAs systems.\\
The Deutsche Forschungsgemeinschaft (Emmy Noether Programm, SFB 463) is acknowledged for financial support.

\end{document}